\documentclass[a4paper,10pt,twoside]{cpc-hepnp}

\usepackage{multicol}
\usepackage{graphicx}
\usepackage{booktabs}
\usepackage{amssymb,bm,mathrsfs,amscd}
\usepackage{bbm}
\usepackage[tbtags]{amsmath}
\usepackage{lastpage}
\usepackage{CJK}
\usepackage{siunitx} 
\usepackage{url}

\begin{document}
\begin{CJK*}{GBK}{song}
\hyphenation{TDHF Skyrme GRAZING Coulomb MNT TLF PLF}

\fancyhead[c]{\small Chinese Physics C~~~Vol. **, No. ** (2018)
**} \fancyfoot[C]{\small **-\thepage}

\footnotetext[0]{Received ** 2018}

\title{Production mechanism of neutron-rich nuclei around $N=126$ in multi-nucleon transfer reaction $^{132}$Sn + $^{208}$Pb\thanks{Supported by  National Natural Science Foundation of China (Projects No.~11705118, No.~11475115 and No.~11647026) and Natural Science Foundation of SZU (grant no. 2016017) }}
\author{%
     Xiang Jiang%
\quad Nan Wang \thanks{Corresponding author} \email{wangnan@szu.edu.cn}%
}
\maketitle
\address{%
College of Physics and Energy,
Shenzhen University, Shenzhen 518060, China
}

\begin{abstract}
Time-dependent Hartree-Fock approach in three dimensions is employed to study the multi-nucleon transfer reaction $^{132}$Sn + $^{208}$Pb at various incident energies above the Coulomb barrier.  Probabilities for different transfer channels are calculated by using particle-number projection method.  The results indicate that neutron stripping (transfer from the projectile to the target) and proton pick-up (transfer from the target to the projectile)  are favored.  Deexcitation of the primary fragments are treated by using the state-of-art statistical code GEMINI++. Primary and final production cross sections of the target-like fragments (with $Z=77$ to $Z=87$) are investigated. The results reveal that  fission decay of heavy nuclei plays an important role  in the deexcitation process of nuclei with $Z>82$.  It is also found that the final production cross sections of neutron-rich (deficient) nuclei  slightly (strongly) depend on the incident energy. 
\end{abstract}



\footnotetext[0]{\hspace*{-3mm}\raisebox{0.3ex}{$\scriptstyle\copyright$}2018
Chinese Physical Society and the Institute of High Energy Physics
of the Chinese Academy of Sciences and the Institute
of Modern Physics of the Chinese Academy of Sciences and IOP Publishing Ltd}%

\begin{multicols}{2}

\section{Introduction}

Neutron-rich nuclei especially those far from $\beta$-stability line are key ingredients to study nuclear structure, elucidate reaction mechanisms and provide information on astrophysical  $r$ process which is responsible for the synthesis of half of the nuclei in nature heavier than iron~\cite{Grawe2007-RPP}. They can be used as projectiles to synthesize super-heavy nuclei located at the island of stability which is one of the goals of the present and future radioactive-ion beam facilities.  

Neutron-rich nuclei in the region of $N=126$ (the last waiting point along the $r$ process) can help to understand the alteration of the shell gap in presence of neutron excess~\cite{Kozulin2012-PRC} and the observed peak structure around $A\sim 195$ in solar $r$-abundance distribution~\cite{Watanabe2015-PRL}. However, due to the difficulties in experiments to produce, detect and identify these nuclei, there are rare relevant investigations on them. Recently, multi-nucleon transfer (MNT) reactions in the vicinity of the Coulomb barrier are evidenced as a feasible route to populate these heavy neutron-rich nuclei. Recent theoretical~\cite{Zagrebaev2007-JPG,Zagrebaev2008-PRL,Zagrebaev2011-PRC,Zhu2017-PLB,Li2018-PLB} and experimental \cite{Kozulin2012-PRC,Beliuskina2014-EPJA,Watanabe2015-PRL,Barrett2015-PRC} results indicate that much larger cross sections can be obtained in MNT reactions than in projectile fragmentations (the main method to produce neutron-rich nuclei with moderate mass). Nevertheless, large-scale experimental performance on this subject is not practical due to the difficulties mentioned above. Therefore, reliable theoretical predictions of probable projectile-target combination, incident energy and so on can help to guide relevant experiments. 

For this purpose, various models are established in recent years. For example,  Langevin type equations of motion~\cite{Zagrebaev2008-PRL,Zagrebaev2007-JPG,Zagrebaev2011-PRC}, semiclassic GRAZING model~\cite{Winther1994-NPA,GRAZING-online},  dinuclear system (DNS) model~\cite{Feng2017-PRC,Zhu2017-PLB,Zhu2017-PRC} and improved quantum molecular dynamics (ImQMD) model~\cite{Wangning2016-PLB,Li2016-PRC,Yao2017-PRC,Li2018-PLB} show success on measurements of MNT in asymmetric reactions.  However, in low-energy reactions around the Coulomb barrier, nuclear structure and quantum effect play very important roles. Therefore, to describe the reactions in nucleonic degrees of freedom quantally and without adjustable parameters is desirable.

Time-dependent Hartree-Fock (TDHF) approach provides a good approximation for microscopically describing the dynamics of such quantum many-body systems at low energies\cite{RevModPhys-75-121}.  It was first proposed by Dirac in 1930 \cite{Dirac-1930}. Present three dimensional (3D) TDHF calculations with modern Skyrme parametrizations  are applied to low-energy heavy-ion collisions on many subjects, for instance, collective vibration~\cite{PhysRevC-68-024302,PhysRevC-71-024301,PhysRevC-71-034314,PhysRevC-71-064328}, fusion reaction~\cite{EPJA-39-243,PhysRevC-88-024617,Jiang2014-PRC2,Jiang2015-EPL}, fission dynamics~\cite{Umar2010-JPG,Goddard2015-PRC,Simenel2014-PRC}, dissipation mechanism~\cite{Maruhn2006-PRC,PhysRevC-86-024608,Dai2014-PRC,Dai2014-SCG,Yu2017-SCG,Guo2018-PLB,Wen2018-PRC}, transfer reaction~\cite{Wangning2016-PLB,Umar2017-PRC}  and so on. Recently, with the help of particle-number projection (PNP) method~\cite{RevModPhys-75-121,Simenel2010-PRL}, the probabilities for different transfer channels can be obtained. To compare with experimental data,  deexcitation of hot primary fragments should be considered. However, this is one of  the drawbacks of dynamical models such as TDHF and ImQMD. One reason is that the time scale of the deexcitation process ($\sim 10^{-17}$~s) is much longer than that of the dynamical collisions.  This can be cured by using the statistical code GEMINI++ after dynamical simulations. Such combination has been widely used recently \cite{Sekizawa2017-PRC,Sekizawa2017-PRCR,Umar2017-PRC,Jiang2014-PRC1,Li2018-PLB}.


This paper is organized as follows. The outlines of the TDHF+GEMINI method and numerical details are given in Section~2. Simulation results are presented and discussed in Section~3. Finally, a summary is enclosed in Section~4.

\section{Formalism and numerical details\label{secII}}

\subsection{TDHF approach and PNP method\label{sec2_1}}

The single-particle wave functions in TDHF  satisfy the fully microscopic equations
\begin{equation}
i \hbar \partial_t \psi_{\alpha} =\hat{h}[\rho] \psi_{\alpha},
\label{eq_TDHF}
\end{equation}
which can be derived from a variational principle \cite{RevModPhys-54-913}, where $\hat{h}$ is the self-consistent mean-field Hamiltonian of single-particle motion and related to the Skyrme functionals, $\rho$ is the one-body density matrix of the independent particle system. The  state of the whole system $\Psi$ is preserved as a Slater determinant which is an anti-symmetrized product of all single-particle wave functions.

In TDHF collisions, nucleons can be exchanged between the reactants once they contact with each other. Single-particle wave functions are partially transfered from the projectile to the target and vice versa. The average proton and neutron numbers of the fragments can be obtained as expectation values of the  particle number operators $\hat{N}^{p}$ (for protons) and $\hat{N}^{n}$ (for neutrons) as $ \langle \Psi | \hat{N}^{q} | \Psi \rangle $ ($q=p,n$).  However, the TDHF states are not eigenstates of the particle number operators but superpositions of these eigenstates. The many-body states can be projected on good particle numbers ($N$ protons or neutrons) by introducing PNP operator expressed as integrals over gauge angles in subspace $V$~\cite{RevModPhys-75-121,Simenel2010-PRL}  
\begin{equation}
\hat{P}_{N;V}=\frac{1}{2\pi}\int_{0}^{2\pi}\mathrm{d}\theta e^{i\theta (\hat{N}_V-N)},
\label{eq_PNPoperator}
\end{equation}
where $\hat{N}_V=\sum_{\alpha=1}^{N_{\textrm{tot}}^{q}} \Theta_V(\bm{r})$ and $ \Theta_V(\bm{r})=1$ if $\bm{r} \in V$ and 0 elsewhere. $N_{\textrm{tot}}^{q}$ is the total states of nucleons with isospin $q$.

At a given incident energy, Eq.~\ref{eq_PNPoperator} can be used to compute the probability $P_{N;V}(b)$ to find $N$ particles in space $V$ for each impact parameter $b$. For later discussions, $P_{N;V}(b)$ is replaced by $P_{N}(b)$ for simplicity. The production cross section of a primary fragment composed of $N$  neutrons and $Z$ protons at a certain incident energy is
\begin{equation}
\sigma_{N,Z}=2\pi \int_{b_{\rm min}}^{b_{\rm max}}  b\mathrm{d}b P_{N,Z}(b),
\label{eq_sigma_pre}
\end{equation}
where $P_{N,Z}(b)=P_N(b) P_Z(b)$, $b_{\rm min}$ is the critical impact parameter inside which fusion happens. In this work, $b_{\rm min}=0$ is used for all incident energies because only binary products are found for all impact parameters. $b_{\rm max}$  is a cutoff impact parameter which depends on the incident energy and  will be given later.  Note that $b_{\rm max}$ should be large enough to make sure that transfer cross sections barely depend on the choice of it.

\subsection{Deexcitation after collisions\label{sec2_2}}

The state-of-art statistical code GEMINI++ is used to treat the deexcitation process of the hot primary fragments produced in TDHF collisions. This code is an improved version of GEMINI which is based on the well-known sequential-binary-decay picture that the individual compound nuclei decay through sequential binary decays of all possible modes (evaporation of neutron, emission of light charged particles, symmetric and asymmetric fission of heavy nuclei), until the resulting products are unable to undergo any further binary decay due to competition with $\gamma$-ray emission~\cite{NPA-483-371}.  The decay width for the evaporation of fragments with $Z\leqslant 2$ is calculated using the Hauser-Feshbach formalism~\cite{Hauser1952}. The transmission coefficient and the level-density parameter are improved in GEMINI++. The fission decay width is calculated using the transition state formalism of Moretto for light system or asymmetric fission of heavy system. Bohr-Wheeler formalism is used in conjunction with the systematics of mass distribution for more symmetric fission of heavy system~\cite{GEMINI}. Further more, extensive comparisons with heavy-ion induced fusion data have been used to optimize the default parameters for GEMINI++.  Further details can be found in Refs.~\cite{NPA-483-371,GEMINI} and references therein. Original version of GEMINI++ with default parameters are used in the present work \cite{PhysRevC-82-014610,PhysRevC-82-044610}.

For a certain primary fragment with  $(N',Z')$, excitation energy $E_{N',Z'}^\ast$ and angular momentum $J_{N',Z'}$ (these quantities will be given in the following subsection), deexcitation process of this fragment should be repeated $M_{\rm trial}$ times due to the statistical nature of GEMINI++. After deexcitation, the number of events in which final fragment with $(N,Z)$ is counted and denoted as $M_{N,Z}$. Then the production cross section  of the final fragment with $(N,Z)$ is given as
\begin{equation}
\begin{aligned}
\sigma_{N,Z}^{\textrm{final}}=& 2\pi \int_{b{\rm min}}^{b_{\rm max}} b\mathrm{d}b \sum_{N' \geqslant N, Z'\geqslant Z}P_{N',Z'}(b)  \frac{M_{N,Z}}{M_{\rm trial}} ,
\end{aligned}
\label{eq_sigma_post} 
\end{equation}

\subsection{Numerical details\label{sec2_3}}

In the present work, 3D unrestricted TDHF code Sky3D \cite{CPC-185-2195} is used to compute the collisions. Skyrme SLy6 parametrization \cite{NPA-635-231} is adopted  for the static and dynamic calculations. The initial nuclei are prepared by the static HF using the damped gradient iteration \cite{NPA-378-418} with imaginary time-step method \cite{NPA-342-111}. Static iterations are performed on $32\times 32\times 32$ Cartesian grids with 1.0 fm grid spacing in all three directions. 

In dynamical calculations, the meshes are extended to $70\times 32\times 70$ while the grid spacing is kept constant as in static HF. The two nuclei are initially placed at a separation distance of 24 fm between their mass centers along the $z-$axis. They are boosted with velocities obtained assuming that they move on a pure Rutherford trajectory with the associated center-of-mass (c.m.) energy $E_{\textup{c.m.}}$ at infinite distance until they reach the initial separation distance. Eq.~\ref{eq_TDHF} is solved iteratively with a time step $\bigtriangleup t=0.2$~fm/c and the exponential propagator is replaced by Taylor series expansion up to order 6. All the reactions are simulated until the separation distance between the primary fragments' mass centers reaches 30~fm. 

In PNP analysis, the integrals over $\theta$ in Eq.~\ref{eq_PNPoperator} are performed with an $M-$point uniform discretion. $M=300$ is adopted for convergence.

To deal with the deexcitation process, the mass and charge numbers, excitation energy and angular momentum of the primary fragment should be provided as inputs of GEMINI++. For a certain transfer channel  with $N$ neutrons and $Z$ protons in the target-like fragment (TLF) while $N_{\rm tot}-N$ neutrons and $Z_{\rm tot}-Z$ protons in the projectile-like fragment (PLF), the mass  and charge numbers of the TLF are $N+Z$ and $Z$, respectively. The total excitation energy  of the system at final distance is adopted as $E_{\rm{tot}}^\ast=E_{\rm{c.m.}} - \rm{TKE}$$+Q_{gg}(N,Z)$ where TKE represents the total kinetic energy of the primary fragments and $Q_{gg}$ is the reaction $Q$ value. The ground-state masses  are adopted from AME2016~\cite{Huang2017-CPC,Wang2017-CPC} and FRDM(2012)~\cite{Moller2016-ADNDT} to calculate the $Q$ value for each transfer channel. $E_{\rm tot}^\ast$ is shared between the  outgoing fragments in proportional to their masses. Average angular momentum of the fragments can be directly obtained in TDHF. The trial event number $M_{\rm trial}$ in GEMINI++ is set as 1000.

\begin{center}
	\centering
	\begin{minipage}{1.0\linewidth}	
		\centering
		\includegraphics[width=0.9\linewidth]{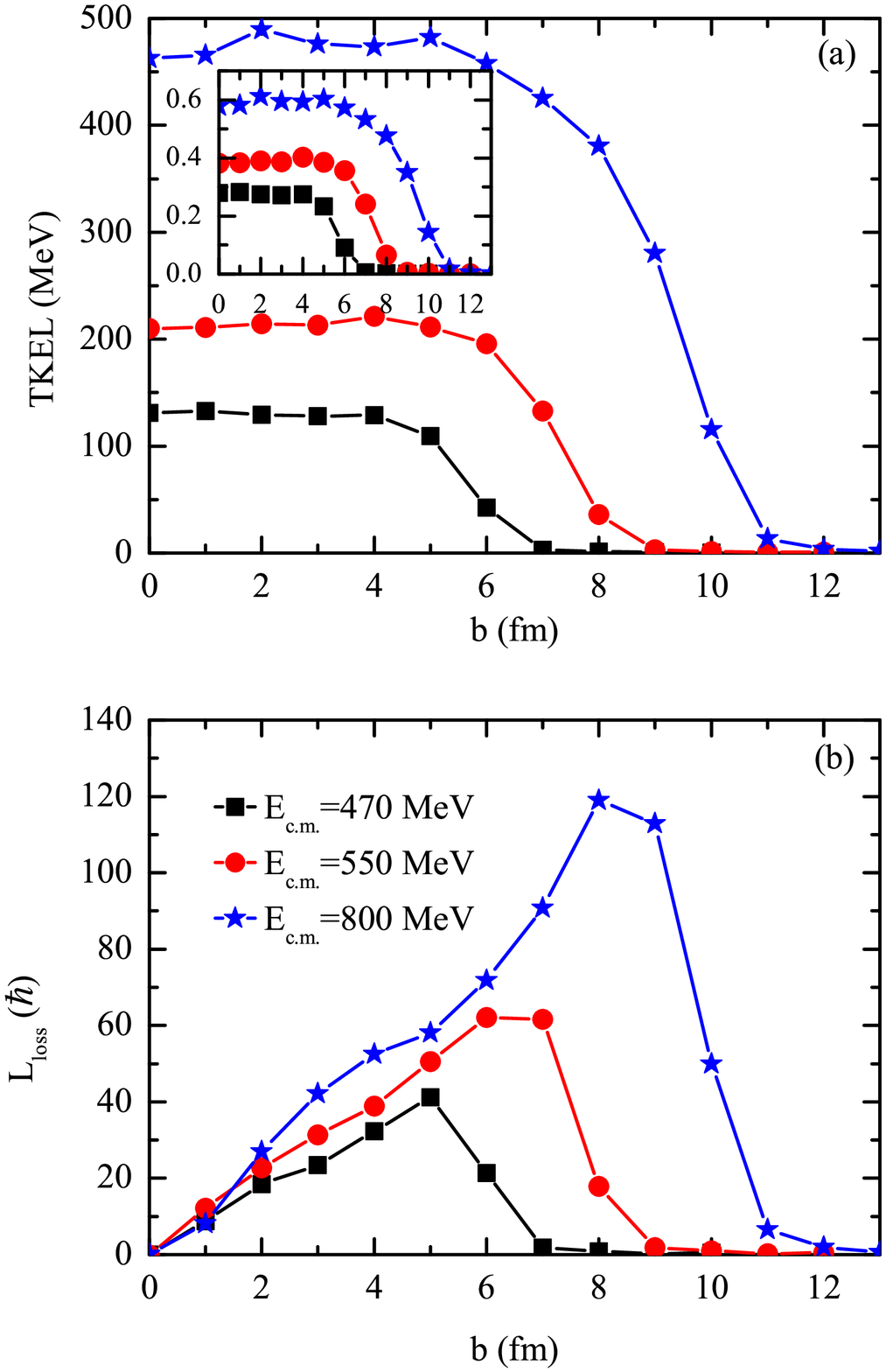}
		\figcaption{(Color online) (a) Total kinetic energy loss  and (b) orbital angular momentum loss as functions of the impact parameter for $^{132}$Sn + $^{208}$Pb at $E_{\rm c.m.}=470$~MeV (black squares), 550~MeV (red circles) and 800~MeV (blue stars), respectively. TKEL/$E_{\rm c.m.}$ as functions of the impact parameter for the three incident energies are shown in the inset of the upper panel. The lines are drawn to guide the eyes.}
		\label{fig_TKEL}
	\end{minipage}
\end{center}

\section{RESULTS AND DISCUSSIONS\label{secIII}}

\subsection{Overview of the TDHF results}\label{sec3_1}

In this subsection, the TDHF results of $^{132}$Sn + $^{208}$Pb without deexcitation are presented. This reaction is one of the candidates for producing heavy neutron-rich nuclei around $N=126$. The simulations are performed at three incident energies, 470, 550 and 800~MeV in the c.m. frame which are about 1.2, 1.4 and 2 times of the Bass barrier ($\sim 400$~MeV). For each bombarding energy, the impact parameter ranges from 0 to $b_{\rm max}$ ($b_{\rm max}=10$~fm, 12~fm and 13~fm for the three energies) with the interval $\Delta b=1$~fm.

\begin{center}
	\centering
	\begin{minipage}{1.0\linewidth}	
		\centering
		\includegraphics[width=0.9\linewidth]{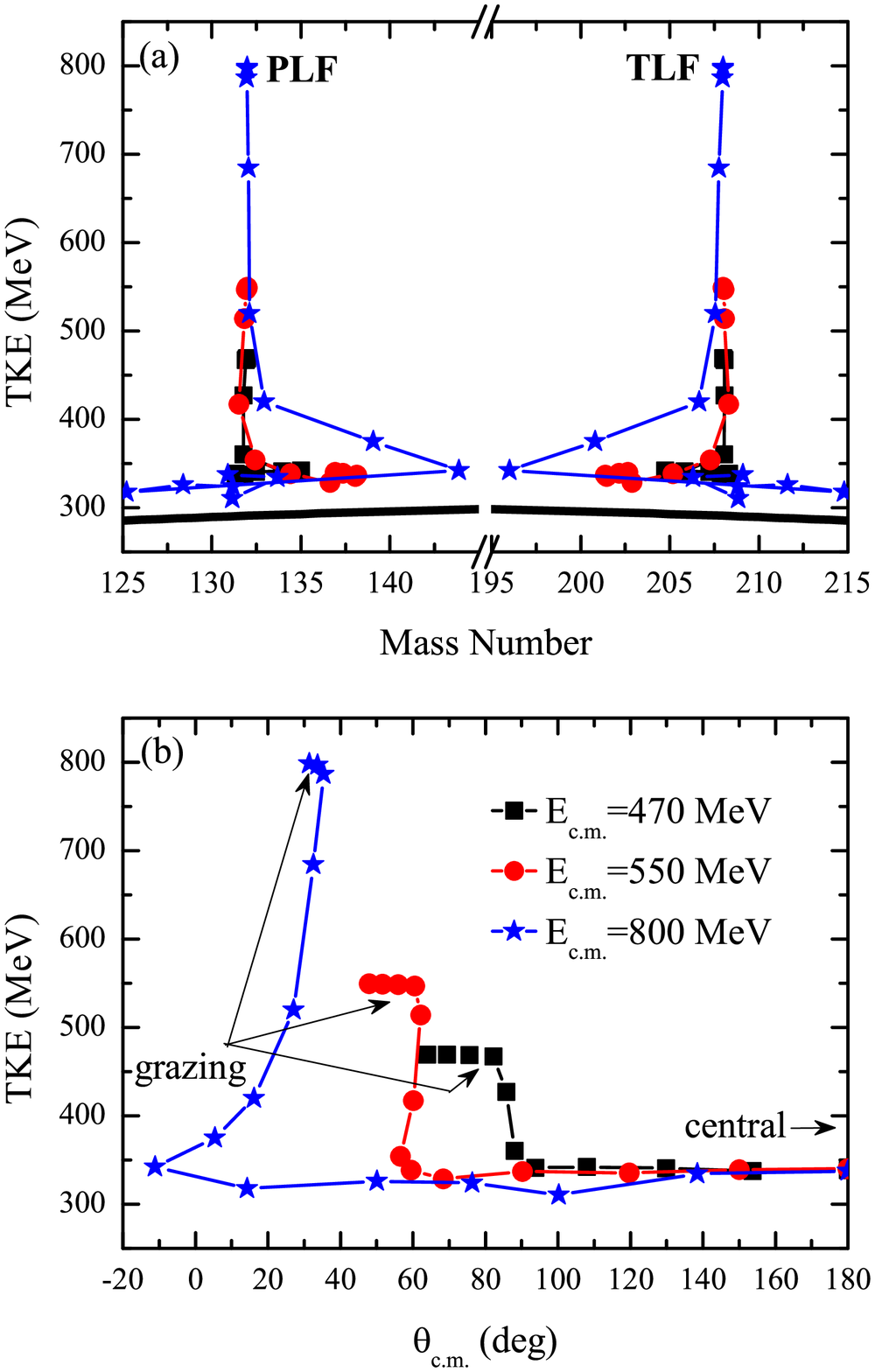}
		\figcaption{(Color online) Total kinetic energy of the outgoing fragments versus (a) masses of the primary fragments  and (b) scattering angle in the c.m. frame for $^{132}$Sn + $^{208}$Pb at $E_{\rm c.m.}=470$~MeV (black squares), 550~MeV (red circles) and 800~MeV (blue stars), respectively. TKE deduced from Viola systematics are shown in the upper pannel as a thick solid line for comparison.}
		\label{fig_TKE}
	\end{minipage}
\end{center}

Shown in Fig.~\ref{fig_TKEL} are the total kinetic energy loss (TKEL) and orbital angular momentum loss ($L_{\rm loss}$) as functions of the impact parameter. TKEL is defined as $E_{\rm c.m.} - \rm TKE$.  $L_{\rm loss}$ is adopted as the difference in initial and final orbital angular momenta.  It can be found in Fig.~\ref{fig_TKEL}(a) that the TKEL shows a smooth dependence on the impact parameter. A plateau pattern can be seen up to $b=5$~fm for all the three energies, and then TKEL decreases gradually with the increasing impact parameter. Strong energy dependence of TKEL and $\rm{TKEL}/E_{\rm c.m.}$ is observed in Fig.~\ref{fig_TKEL}(a) and the inset.  They both increase with the increasing incident energy. This is due to the fact that more nucleons are exchanged between the projectile and the target nuclei at higher energies. As a consequence, the collisions become more violent and more energies of collective motion should be dissipated. We note in Fig.~\ref{fig_TKEL}(b) that $L_{\rm loss}$ increases linearly with the impact parameter up to $b=5$, 7 and 8~fm for the three incident energies, respectively. At larger impact parameters, it decreases gradually as expected. Energy dependence is also observed for  $L_{\rm loss}$.

\begin{center}
	\centering
	\begin{minipage}{1.0\linewidth}	
		\centering
		\includegraphics[width=0.9\linewidth]{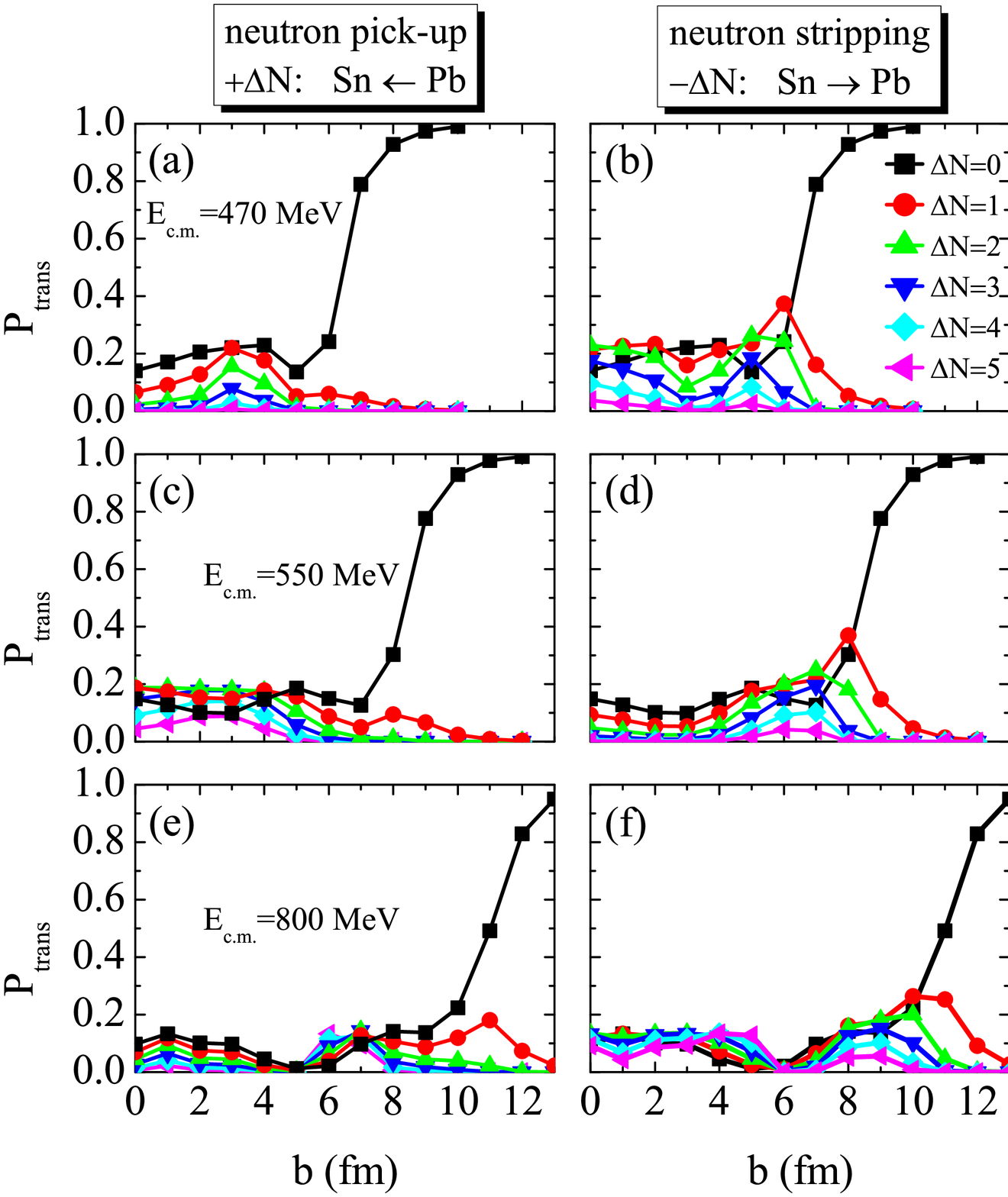}
		\figcaption{(Color online) Probabilities of  neutron pick-up (from the target to the projectile, left panels) and neutron stripping (from the projectile to the target, right panels)  channels as functions of the impact parameter for $^{132}$Sn + $^{208}$Pb at $E_{\rm c.m.}=470$~MeV (top panels), 550~MeV (middle panels) and 800~MeV (bottom panels), respectively.}
		\label{fig_PtransN}
	\end{minipage}
\end{center}

Fig.~\ref{fig_TKE}(a) shows TKE as a function of the masses of the primary fragments. The results of Viola systematics~\cite{Viola1985-PRC,PhysRevC-45-1229} (the relative momentum in the entrance channel is fully damped and TKE originates from Coulomb repulsion of the outgoing fragments at a scission configuration) are also shown as a comparison. Two very prominent peaks around the initial masses of the reactants with almost no kinetic energy loss are found in TDHF results, which can be interpreted as the results of (quasi)elastic scattering at peripheral collisions. At small TKE, which corresponds to more central collisions, we note that TKE is kept around 340~MeV for all the three energies in a wide mass range. We note that this value is about 30~MeV above the Viola systematics which reflects properties of deep-inelastic collisions. Moreover, the mass distribution gets wider with the increasing incident energy. Shown in Fig.~\ref{fig_TKE}(b) is the TKE versus the scattering angle in c.m. frame ($\theta_{\rm c.m.}$).  We note that when TKE is around 340~MeV,  $\theta_{\rm c.m.}$ spreads in a very wide range which also shows deep-inelastic characteristics. For $b=6$~fm at $E_{\rm c.m.}=800$~MeV, $\theta_{\rm c.m.} \approx -\ang{20}$ is observed which is a result of the strong competition of nuclear attraction, Coulomb repulsion and centrifugal effect. At grazing regions, $\theta_{\rm c.m.}$ is almost kept  constant around $\ang{85}$, $\ang{60}$ and  $\ang{30}$ for $E_{\rm c.m.}=470$, 550 and 800~MeV, respectively.

\begin{center}
	\centering
	\begin{minipage}{1.0\linewidth}
		\centering		
		\includegraphics[width=0.9\linewidth]{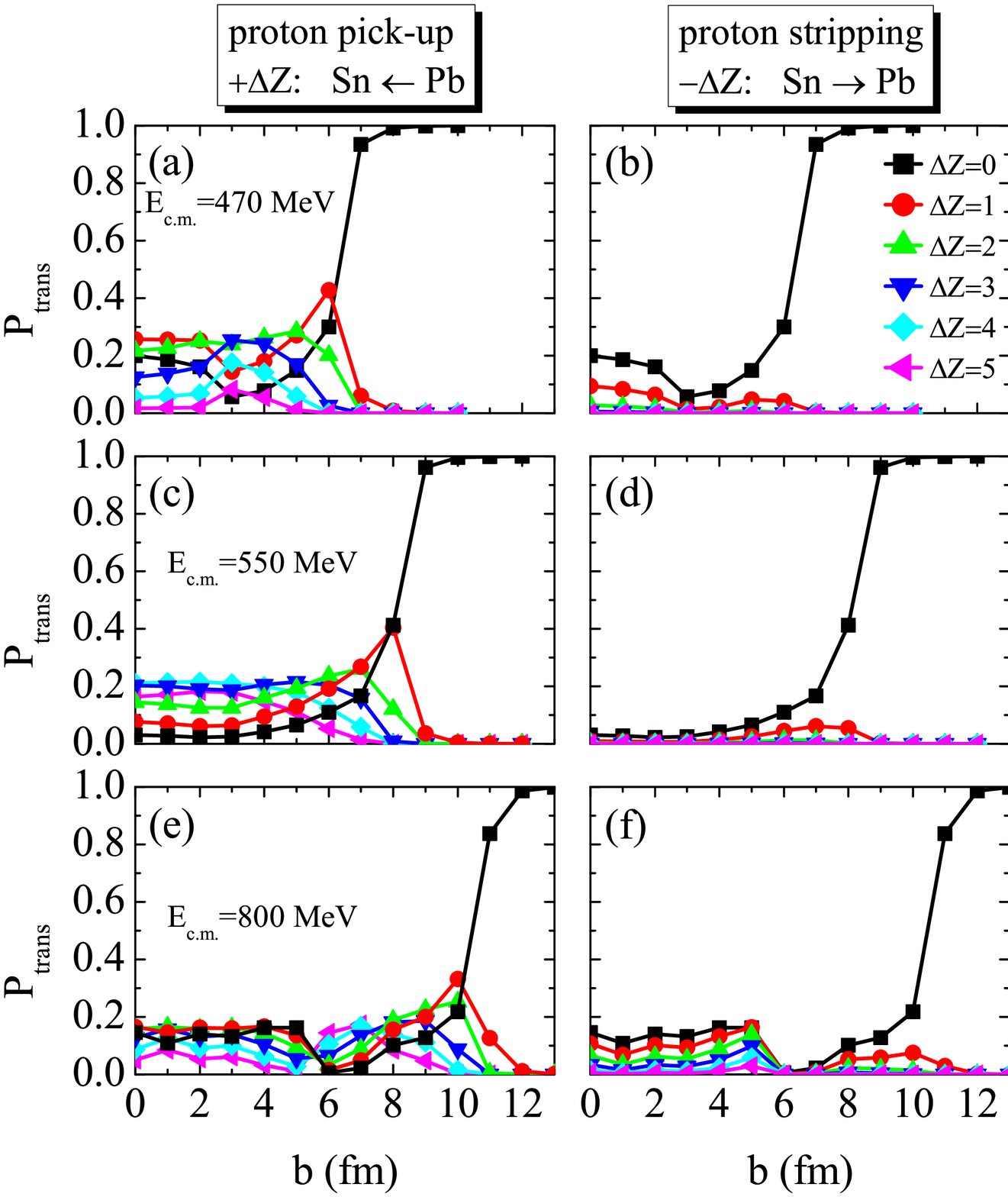}
		\figcaption{(Color online) Same as Fig.\ref{fig_PtransN}, but for proton transfer channels.}		
		\label{fig_PtransZ}	
	\end{minipage}
	
\end{center}

The  probabilities for different nucleon transfer channels at each impact parameter are obtained from the TDHF final wave functions by using PNP method. The results of neutron pick-up (from the target to the projectile) and stripping (from the projectile to the target) channels  are shown in Fig.~\ref{fig_PtransN} while those of the proton pick-up and stripping channels are presented in Fig.~\ref{fig_PtransZ}, respectively.  Some gross features can be seen from Figs.~\ref{fig_PtransN} and \ref{fig_PtransZ} that probabilities of $\Delta N=0$ or $\Delta Z=0$ increase with the increasing impact parameter in semi-peripheral to peripheral regions. Nucleon transfer process is much reduced in peripheral collisions for all the energies. This is because at peripheral regions, the projectile and target collide gently and the interacting time is not long enough for multi-nucleon exchange.   In more central collisions, the probabilities show complicate dependence on the impact parameter and incident energy.

\begin{center}
	\centering
	\begin{minipage}{1.0\linewidth}
		\centering		
		\includegraphics[width=0.75\linewidth]{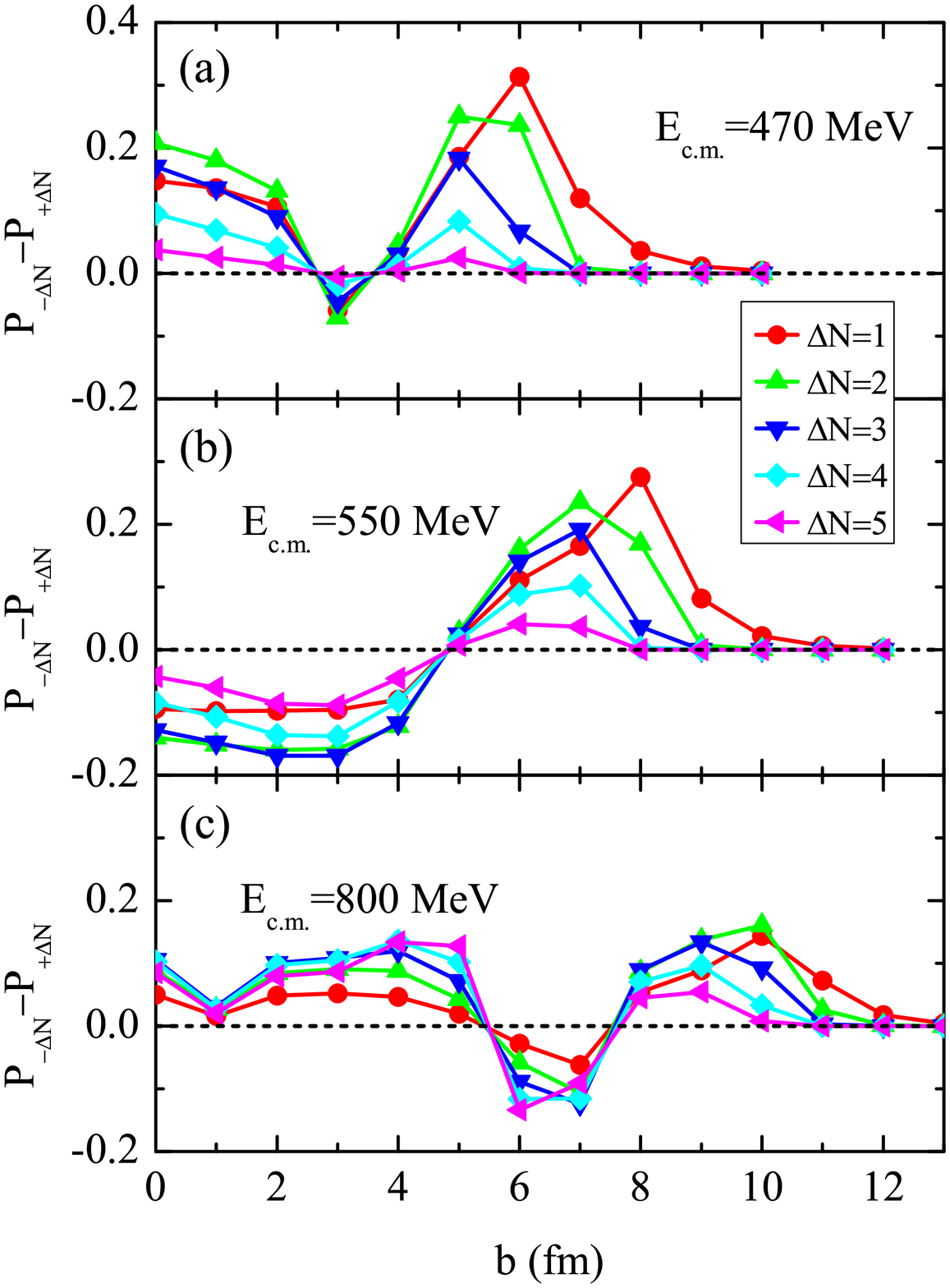}
		\figcaption{(Color online) Differences between the probabilities of neutron stripping ($-\Delta N$) and pick-up ($+\Delta N$) channels as functions of the impact parameter for $^{132}$Sn + $^{208}$Pb at $E_{\rm c.m.}=470$~MeV (top panels), 550~MeV (middle panels) and 800~MeV (bottom panels), respectively.}
		\label{fig_PtransNdiff}
	\end{minipage}	
\end{center}

\begin{center}
	\centering
	\begin{minipage}{1.0\linewidth}
		\centering		
		\includegraphics[width=0.75\linewidth]{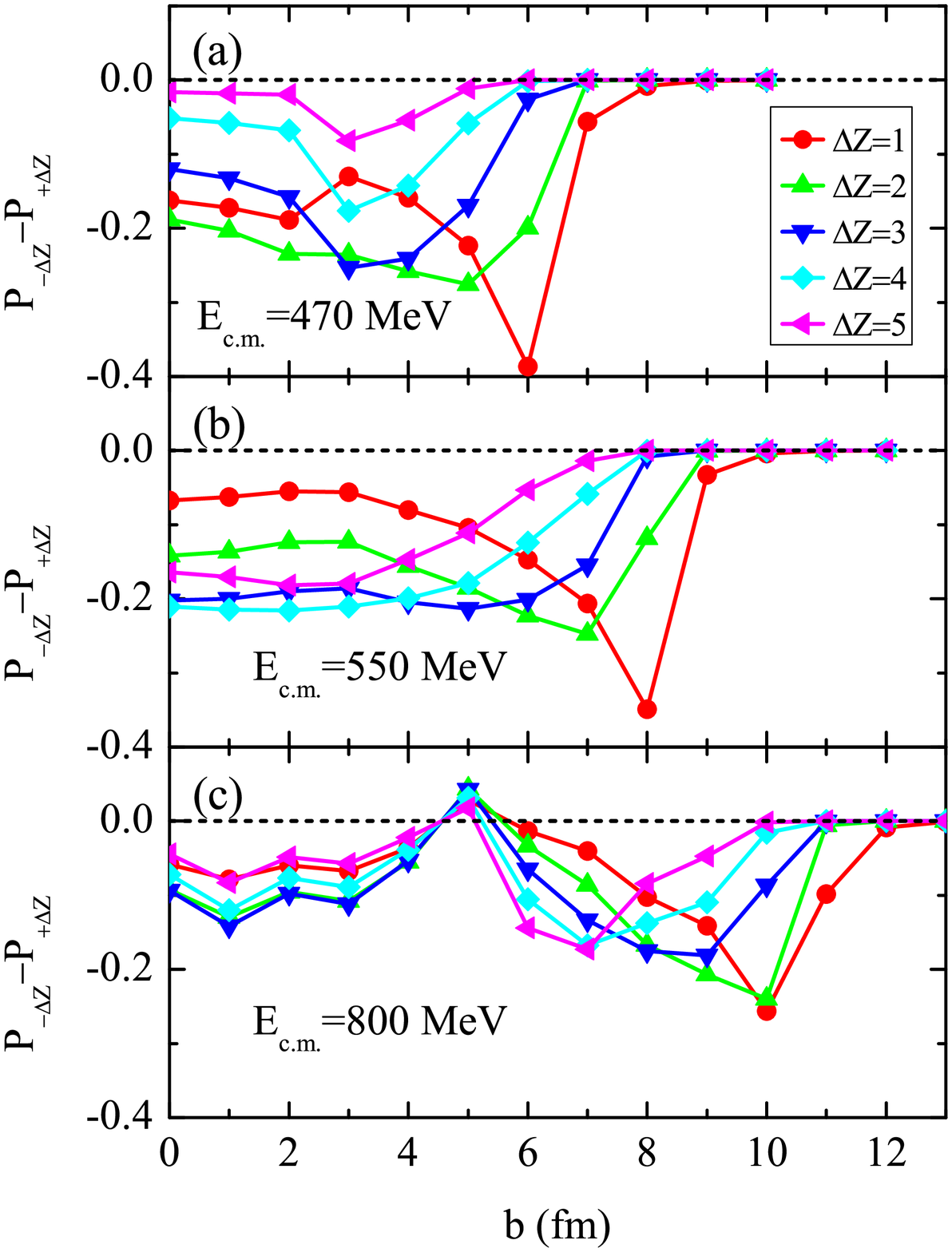}
		\figcaption{(Color online) Same as Fig.\ref{fig_PtransNdiff}, but for proton transfer channels.}		
		\label{fig_PtransZdiff}	
	\end{minipage}
	
\end{center}

To get a clearer insight, the differences between the probabilities of neutron stripping ($-\Delta N$) and pick-up ($+\Delta N$) channels (denoted as $P_{-\Delta N}-P_{+\Delta N}$) are shown in Fig.~\ref{fig_PtransNdiff}. One can see that at $E_{\rm c.m.}=470$~MeV,  probabilities of neutron stripping channels are larger than those of the pick-up channels for all impact parameters except for $b=3$~fm. At $E_{\rm c.m.}=550$~MeV, probabilities of neutron stripping channels are smaller  than those of the pick-up channels for $b\leqslant 4$~fm while larger for $b\geqslant 5$~fm.  Probabilities of neutron stripping channels are found larger than those of the pick-up channels for all impact parameters except for $b=6$ and 7~fm at $E_{\rm c.m.}=800$~MeV. Though neutron pick-up dominates at certain impact parameters (generally in central and semi-central regions), the total production cross sections for neutron stripping channels are still larger than those for neutron pick-up channels because large impact parameter has larger contributions to the total cross sections as seen from Eq.~\ref{eq_sigma_pre}. The above results indicate that neutron transfer from the projectile to the target is favored.  

The differences between the probabilities of proton stripping ($-\Delta Z$) and pick-up ($+\Delta Z$) channels (denoted as $P_{-\Delta Z}-P_{+\Delta Z}$) are shown in Fig.~\ref{fig_PtransZdiff}. One can find that  proton transfer from the target to the projectile is favored for almost all cases. Such nucleon transfer modes are quite beneficial for producing neutron-rich nuclei.

\subsection{Primary and final production cross sections}\label{sec3_2}
In this subsection, the  production cross sections of the primary TLFs obtained  in THDF and those of the final states after deexcitation are calculated by Eqs.\ref{eq_sigma_pre} and \ref{eq_sigma_post}.

Shown in Fig.~\ref{fig_isotopes} are the isotopic production cross sections of different proton-transfer channels from $Z=77$ to $Z=87$. One can find that for each channel, the primary isotopes distribute in a broader range comparing with the final ones. The primary isotopic distributions also get broader with the increasing incident energy. The largest cross sections for isotopes with $Z<82$ depend weakly on the incident energy. For isotopes with $Z>82$, however, the largest cross sections of them show strong dependence on the incident energy. For example, the largest cross section of  francium at $E_{\rm c.m.}=800$~MeV is larger than those of the two lower-energy cases by  2 to 3 orders of magnitude.

\end{multicols}
\vspace{5mm}
\ruleup
\begin{center}
	\centering
	\includegraphics[width=16cm]{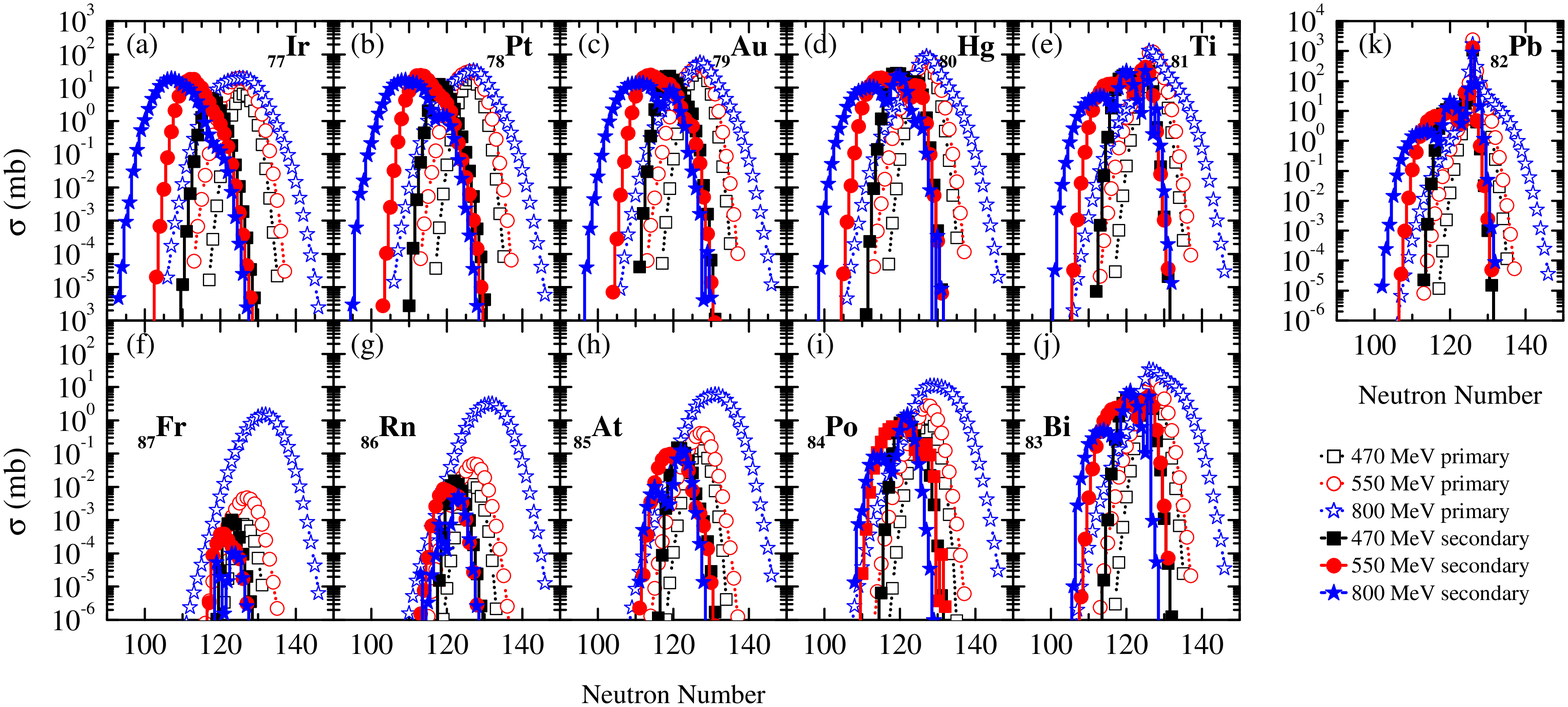}
	\figcaption{(Color online) Primary (empty symbols with dashed lines) and final (solid symbols with solid histograms) production cross sections of the TLFs for $^{132}$Sn + $^{208}$Pb at $E_{\rm c.m.}=470$~MeV (black squares), 550~MeV (red circles) and 800~MeV (blue stars), respectively.}
	\label{fig_isotopes}
\end{center}
\vspace{-5mm}
\ruledown
\vspace{5mm}
\begin{multicols}{2}

After deexcitation, prominent differences between the primary and final results can be found. Firstly one can see that the distributions for all channels move toward lower $N$. Exotic nuclei with large isospin asymmetry can not survive in the deexcitation process. This can be interpreted as the results of neutron evaporation in the deexcitation process. Secondly, prominent decreases for the peak values of the final cross sections are observed for $Z\geqslant 79$, especially for higher incident energies. 

To get a deeper insight on these results, the decay modes of a primary fragment with $(N,Z)$ are divided into three types: neutron evaporation (no fission and without light charged particle emission, denoted as mode 1), light charged particle emission (accompanied with neutron evaporation but without fission, denoted as mode 2)  and fission of heavy nuclei (accompanied with neutron evaporation and light charged particle emission, denoted as mode 3). The numbers of events for each binary decay mode in GEMINI++ are counted and denoted as $M_{\textrm{mode}~i}$, where $i=1, 2, 3$. The relative ratios of these modes are calculated by $\eta_i=\frac{\sigma_{\textrm{mode}~ i}}{\sigma_Z} \times 100\%$,  where $\sigma_{\textrm{mode}~i}=2\pi \sum_N \int_{b_{\textrm{min}}}^{b_{\textrm{max}}}b\textrm{d}b P_{N,Z}(b) \frac{M_{\textrm{mode}~i}}{M_{\rm trial}}$ and  $\sigma_Z=\sum_{N} \sigma_{N,Z}$. 

\vspace{2mm}
		
The results of $Z=77$ and 87 are shown as examples. 
At $E_{\rm c.m.}=470$~MeV, $\eta_1=99.88\%$, $\eta_2=0.1\%$ and $\eta_3=0.02\%$ are obtained for  $Z=77$, while the values are 70\%, 0.8\% and 29.2\% for Z=87. At $E_{\rm c.m.}=550$~MeV, the contributions are about 87.3\%, 12.3\% and 0.4\% for Z=77, while 10.4\%, 11.9\% and 77.7\% for Z=87. At $E_{\rm c.m.}=800$~MeV, the contributions are about 16.7\%, 75.9\% and 7.4\% for Z=77, while 0\%, 27.8\% and 72.2\% for Z=87.  Thus we can conclude that fission decay plays an important role for the deexcitation of proton stripping channels. 

In Fig.~\ref{fig_isotopes} (e), (j) and (k), peaks around $N=126$ are seen. This is the result of quantum effect of $N=126$ closed neutron shell. We also note that the final production cross sections of  neutron-rich isotopes show weak dependence on the incident energy. In contrast, those of neutron-deficient isotopes depends strongly on the incident energy. This phenomenon was also observed in Ref.~\cite{Feng2017-PRC}.

The production cross sections of  neutron-rich heavy nuclei with $N=126$ are extracted from Fig.~\ref{fig_isotopes} and plotted in Fig.~\ref{fig_N=126}. Both the primary and final cross sections are given.

\begin{center}
	\centering
	\begin{minipage}{1.0\linewidth}	
	\centering
	\includegraphics[width=0.9\linewidth]{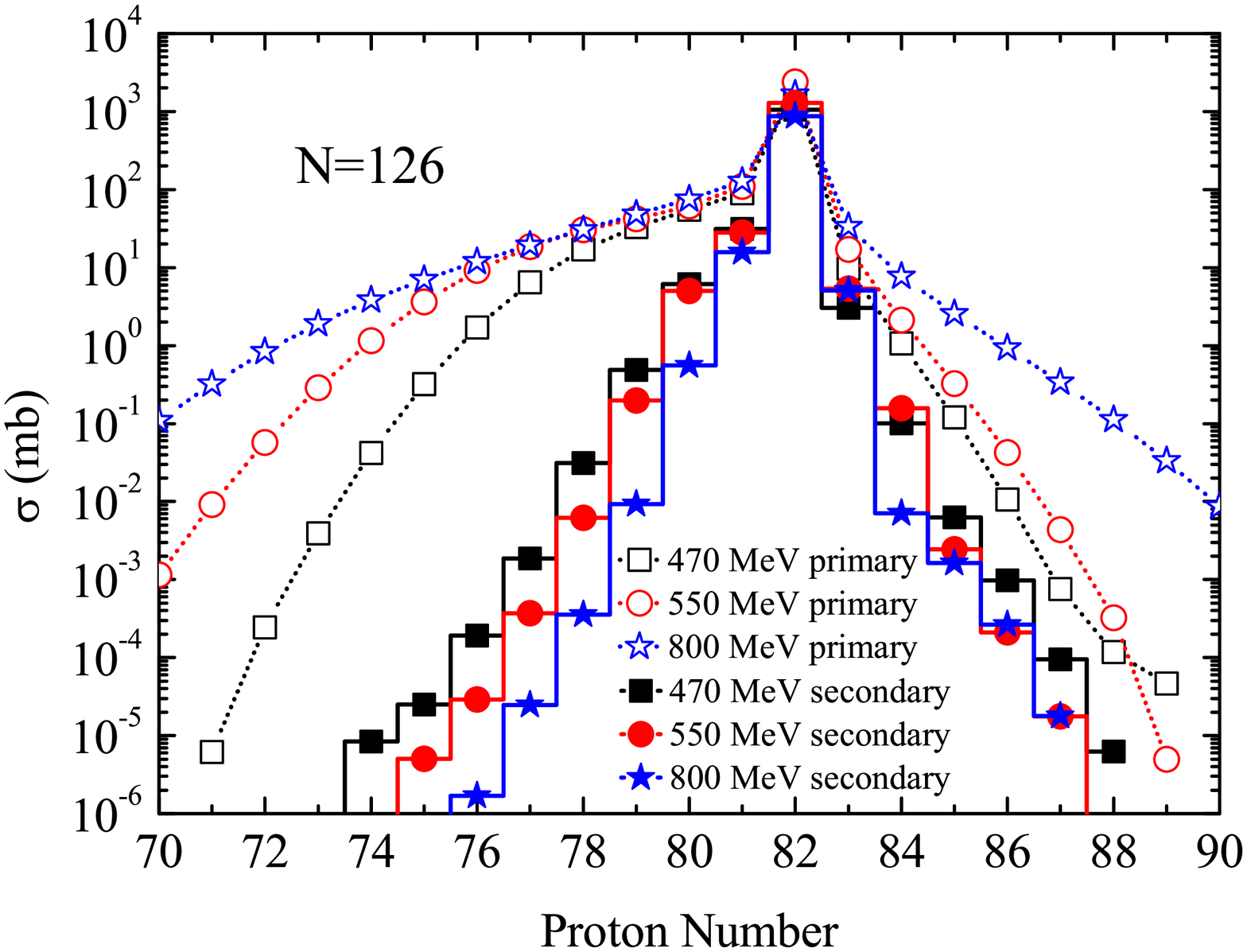}
	\figcaption{(Color online) Primary (empty symbols with dashed lines) and final (solid symbols with solid histograms) isotonic production cross sections of the nuclei with $N=126$ for  $^{132}$Sn + $^{208}$Pb at $E_{\rm c.m.}=470$~MeV (black squares), 550~MeV (red circles) and 800~MeV (blue stars), respectively. }
	\label{fig_N=126}
\end{minipage}
\end{center}

From Fig.~\ref{fig_N=126} it can be found that both the primary and final products distribute in a broad range with the peak centering at $Z=82$. From above discussions on Figs. 3 and 4, it is clear that $^{208}$Pb nuclei are mainly from peripheral collisions which correspond to (quasi)elastic scattering. The excitation energies of these fragments are small (seen from Fig. 1) and most of them can survive in the deexcitation process. So both the primary and final cross sections of $^{208}$Pb are mainly attributed to (quasi)elastic channels. One should note that these cross sections depend on $b_{\textrm{max}}$ (in this work $b_{\textrm{max}}$=10, 12 and 13 fm are used for $E_{\rm c.m.}$=470, 550 and 800 MeV, respectively. As we mentioned at the end of  Section~2.1, $b_{\textrm{max}}$ should be large enough to make sure that the transfer cross sections barely depend on the choice of it). Unfortunately, it's very hard to separate the total cross sections into components of ``(quasi)elastic scattering" or ``transfer channels" which makes the cross sections of $^{208}$Pb less meaningful. This problem is also found for  $^{208}$Pb in Fig.~\ref{fig_isotopes}(k). But it will not change the conclusions of the present work. 

We also note that the primary production cross sections increase with the incident energy, especially for large $|\Delta Z|$ regions. For example, at $E_{\rm c.m.}=800$~MeV, the cross sections for $Z=71$ are orders of magnitude larger than those of the two lower energies.  This is because more nucleons are exchanged as the incident energy increases. After deexcitation, however, the production cross sections decrease rapidly as $|\Delta Z|$ increases. Meanwhile, they decrease with the increasing incident energy, especially for the lower-$Z$ regions.  The cross sections of 800~MeV are at least one order of magnitude  lower than those of 470~MeV. This is because much higher excitation energies are involved at $E_{\rm c.m.}=800$~MeV and most of those nuclei decay in the deexcitation process.

\section{Summary\label{secIV}}

We apply 3D TDHF with PNP method to MNT reaction  $^{132}$Sn + $^{208}$Pb at various incident energies  above the Coulomb barrier. Collision dynamics show characteristics of deep-inelastic collisions. Impact parameter dependence of the probabilities for different transfer channels indicate that neutron stripping and  proton pick-up are favored. Such transfer modes benefit the production of neutron-rich nuclei for the TLF. Production cross sections of the primary fragments depend strongly on the incident energy. 

The deexcitation of the primary fragments formed in TDHF are treated by using the state-of-art statistical code GEMINI++. Final isotopic production cross sections of the TLF from $Z=77$ to $Z=87$ are investigated. The results reveal that  fission decay of heavy nuclei plays an important role for proton stripping channels. Shell effect is found to be important in the deexcitation for few-nucleon-transfer channels. The production cross sections of  neutron-rich nuclei show slight dependence on the incident energy, even at energy up to 2 times of the Bass barrier. The results of neutron-deficient nuclei, however, depend strongly on the incident energy that much larger cross sections of neutron-deficient nuclei are obtained at higher energies. Finally, production cross sections of neutron-rich nuclei with $N=126$ are calculated. They decrease gradually as the incident energy increases, especially for lower-$Z$ isotones. This is due to larger excitation energies are involved  in higher-energy collisions. 

This work presents a self-consistent method to predict yields of neutron-rich nuclei in MNT transfer reactions. However, due to the lack of nucleon-nucleon correlations, mean-field models can not describe large fluctuations well and the width of distributions are underestimated. As a first step, pairing effect can be included in TDHF for transfer reactions. Relevant work is in progress.

\acknowledgments{We thank  Prof. H. Z. Liang for helpful discussions.}

\end{multicols}

\vspace{10mm}

\begin{multicols}{2}

\vspace{3mm}



\end{multicols}

\clearpage

\end{CJK*}

\end{document}